\documentclass[10pt
]{article}
\usepackage[a4paper]{geometry}
\usepackage{graphicx}
\usepackage{amsmath,amssymb}
\usepackage{authblk}
\usepackage{cite}
\usepackage{subcaption}
\usepackage{xfrac}
\usepackage{authblk}
\usepackage{booktabs}

\usepackage{tikz}
\usetikzlibrary{decorations.pathmorphing}
\usetikzlibrary{decorations.markings}
\usetikzlibrary{shadows}
\usepackage{pgfplots}

\usepackage[abs]{overpic}

\usepackage[sc]{mathpazo}
\usepackage[T1]{fontenc}

\newcommand\eq[1] {(\ref{#1})}

\newcommand{\beqa}{\begin{eqnarray}}
\newcommand{\eeqa}[1]{\label{#1}\end{eqnarray}}
\newcommand{\bequ}{\begin{equation}}
\newcommand{\eequ}[1]{\label{#1}\end{equation}}

\newcommand{\Gb}{\beta}

\newcommand{\Go}{\omega}

\newcommand{\GD}{\Delta}

\newcommand{\CF}{{\cal F}}

\newcommand{\beq}{\begin{equation}}
\newcommand{\eeq}{\end{equation}}
\newcommand{\overliner}{\begin{eqnarray}}
\newcommand{\earr}{\end{eqnarray}}
\newcommand{\beqn}{\begin{equation*}}
\newcommand{\eeqn}{\end{equation*}}
\newcommand{\overlinern}{\begin{eqnarray*}}
\newcommand{\earrn}{\end{eqnarray*}}
\newcommand{\prt}{\partial}

\newcommand{\fr}{\frac}

\usepackage[defaultsans]{droidsans}
\usepackage[eulergreek]{sansmath}
\pgfplotsset{
  tick label style = {font=\sansmath\sffamily},
  every axis label = {font=\sansmath\sffamily},
  legend style = {font=\sansmath\sffamily},
  label style = {font=\sansmath\sffamily}
}

\renewcommand{\vec}[1]{{\mathbf{#1}}}
\newcommand{\vect}[1]{{\boldsymbol{#1}}}

\title{Transformation cloaking and radial approximations for flexural waves in elastic plates}

\author[1,2]{M. Brun\thanks{Author for correspondence: mbrun@unica.it}}
\author[3]{D.J. Colquitt}
\author[4]{I.S. Jones}
\author[2]{A.B. Movchan}
\author[2]{N.V. Movchan}

\affil[1]{\small Dipartimento di Ingegneria Meccanica, Chimica e dei Materiali, Universit\'{a} di Cagliari, Piazza d'Armi, I-09123 Cagliari, Italy}
\affil[2]{Department of Mathematical Sciences, University of Liverpool, Liverpool, L69 3BX, U.K.}
\affil[3]{Department of Mathematics, Imperial College London, South Kensington, London, SW7 2AZ, U.K.}
\affil[4]{School of Engineering, John Moores University, Liverpool, L3 3AF, U.K.}

\date{\today}

\begin{document}

\maketitle

\begin{abstract}
It is known that design of elastic cloaks is much more challenging than the design idea for acoustic cloaks, cloaks of electromagnetic waves or scalar problems of anti-plane shear.
In this paper, we address fully the fourth-order problem and develop a model of a broadband invisibility cloak for channelling flexural waves in thin plates around finite inclusions. We also discuss an option to employ efficiently an  elastic pre-stress and body forces to achieve such a result.   An asymptotic derivation provides a rigorous link between the model in question and elastic wave propagation in thin solids. This is discussed in detail to show connection with non-symmetric formulations in vector elasticity studied in
earlier work.  
\end{abstract}

\begin{center}
\end{center}

\section{Introduction}

There is a theoretical and practical interest in wave cloaking in the context of metamaterials, as outlined in the publications  \cite{pendry2006,schurig2006,leonhardt2006,milton2006,chen2007,cummer2007,norris2011,farhat2008PRL}.
Dynamic effects include anisotropy and localization \cite{ColJonMovMov2011,CarBruMovMovJon2014}, which can be be interpreted in the context of the homogenization theory.
In this regard, we would like to refer to the work \cite{LiuChanSheng2005,MeiLiuWenSheng2007a,ShengMeiLiuWen2007b}  addressing the notion of an effective dynamic mass density in structured composites and acoustic materials, as well as analytical studies of dynamic localization in phononic crystals. The approach of the dynamic homogenization has been systematically applied in \cite{CraKapPos2010,NolCraKap2011} to vibrations of inertial lattice systems.  
The idea of a so-called ``geometric optics'' transformation leading to a radially symmetric ``push-out'' cloak, is 
commonly used for computational and experimental implementation \cite{pendry2006,brun2009,chen2010,stenger2012,ColBruGeiMovMovJon2014}. In scalar problems, where the governing equation is reduced to Helmholtz form, such a transformation proves to be extremely efficient, leading to a model of a specially designed highly anisotropic inhomogeneous material occupying the cylindrical cloaking layer and channeling incident waves around a finite scatterer (an inclusion or a void). The continuum model of an invisibility cloak leads to singular behaviour of the theoretical material at the inner boundary of the cloaking region adjacent to the scatterer. In a practical implementation, a continuum cloak is replaced by a micro-structured composite, and examples of such implementation include water waves \cite{farhat2008PRL}, flexural plates \cite{stenger2012} and  acoustics \cite{Zigoneau2014}. This micro-structure makes the cloak approximate, and such an approximation is frequency sensitive.  A special challenge is presented for vector problems of elasticity discussed in \cite{milton2006,brun2009,norris2011,norris2012}.

The present paper addresses cloaking for flexural waves in Kirchhoff elastic plates. Firstly, we show that the governing equations are not invariant with respect to the radial ``push-out'' transformation \cite{GreLasUhl2003,leonhardt2006}.
This observation implies that the cloaking design procedure, well developed for acoustics, vibration of elastic membranes  and anti-plane shear problems (see, for example,
\cite{norris2008,
ColJonMovMovBruMcP2013}), does not apply to problems of flexural vibrations of elastic plates. Elastic Kirchhoff plates possess flexural rigidity and their out-of-plane vibrations are governed by a fourth-order partial differential equation. One of the main challenges appears to be the presence of propagating and evanescent waves representing solutions of the Helmholtz and modified Helmholtz equations, and the coupling of such waves via the boundary and interface contact conditions.
In numerical simulations, it is apparent that in many configurations the flexural waves are led by their Helmholtz component (see, for example,
\cite{McPMovMov2009, 
AntCra2012}). However, for cloaking problems the multi-scale nature of a metamaterial makes the problem more challenging and it is not apparent that such decoupling is possible.    

There is strong experimental evidence, as published in
\cite{stenger2012} and also outlined in
\cite{McPMov2012}, that within a predefined frequency range a by-pass system can be implemented around a finite obstacle in a flexural Kirchhoff plate. Such a by-pass system is evidently an approximate cloak, that would benefit strongly from an in-depth analysis paving the way to a broadening of the frequency range for the cloaking effect.  

We explain the derivation of such an approximate cloaking model and present illustrative numerical examples which agree with the experimental evidence \cite{stenger2012}.

The paper
\cite{ColBruGeiMovMovJon2014} has shown, for a model of a square cloak, that a formulation for flexural waves in a Kirchhoff plate, after the cloaking transformation,  includes additional terms in the governing equation; these may represent in-plane body forces and pre-stress. This approach provides a consistent procedure justifying the additional terms in the governing equation and cloaking is effective across the whole frequency range admissible for the plate model.
Motivated by results of
\cite{ColBruGeiMovMovJon2014}, we also develop the full cloaking model for the radial ``push-out'' transformation, and obtain explicit closed form representation for the pre-stress required to have a broadband cloak for flexural waves.

Finally, we present a detailed asymptotic analysis, which establishes a connection between the transformed equations for the fourth-order model of flexural waves and those for a vector problem of elasticity in thin solids.

\section{Application of the radial ``push-out'' transformation to a Kirchhoff-Love plate}
\label{sec:trans-plates}

We begin with a simple case of the equation governing the out-of-plane displacement amplitude $w(\vec{X})$ of an orthotropic homogeneous  plate, in the absence of applied in-plane forces, under pure bending. As in \cite{leissa1969}, the fourth-order partial differential equation is 
\begin{multline}
D_R\frac{\partial^4 w}{\partial R^4}+\frac{2}{R^2}D_{R \Theta}\frac{\partial^4 w}{\partial R^2 \partial \Theta^2}+D_{\Theta}\frac{1}{R^4}\frac{\partial^4 w}{\partial \Theta^4}+\frac{2}{R}D_R\frac{\partial^3 w}{\partial R^3}- \frac{2}{R^3}D_{R \Theta}\frac{\partial^3 w}{\partial R \partial \Theta^2}-\frac{1}{R^2}D_\Theta\frac{\partial^2 w}{\partial R^2 }\\
+\frac{2}{R^4}(D_\Theta+D_{R \Theta})\frac{\partial^2 w}{ \partial \Theta^2}+\frac{1}{R^3}D_\Theta\frac{\partial w}{\partial R}- 
{\rho h}
\omega^2w=0,
\label{eq:untransformed-plate-eq}
\end{multline}
where $D_R$, $D_\Theta$ and $D_{R\Theta}$ are the flexural rigidities, $\rho$ and $h$ are the mass density per unit volume and thickness of the plate, respectively, and $\omega$ is the angular frequency.

The constitutive relations that define the moments are
\begin{eqnarray}
\nonumber
M_R = - D_R \left[ \frac{\prt^2 w}{\prt R^2} + \nu_\Theta \left( \fr{1}{R} \fr{\prt w}{\prt R} + \fr{1}{R^2} \fr{\prt^2 w}{\prt \Theta^2}\right) \right],\\
\nonumber
M_\Theta = - D_\Theta \left( \fr{1}{R} \fr{\prt w}{\prt R} +\fr{1}{R^2} \fr{\prt^2 w}{\prt \Theta^2} + \nu_R \fr{\prt^2 w}{\prt R^2}\right), \\
M_{R\Theta} = - 2 D_K \fr{\prt }{\prt R} \left(  \fr{1}{R} \fr{\prt w}{\prt \Theta} \right),
\label{eqnplate001}
\end{eqnarray}
where $\nu_R$ and $\nu_\Theta$ are the values of the Poisson's ratios in the radial and tangential directions respectively. We also note that $D_K = \fr{1}{2} (D_{R \Theta} - D_R \nu_\Theta)$, and the rigidities $D_R$ and $D_\Theta$ satisfy the following symmetry relation 
\begin{equation}
D_R \nu_\Theta = D_\Theta \nu_R. \label{eq:untransformed-plate-eq-a}
\end{equation}

Further, if  the plate is isotropic and homogeneous, then the equation \eqref{eq:untransformed-plate-eq} will have $D_R=D_\Theta=D_{R\Theta}=D^{(0)}$ where $D^{(0)}$ is the flexural rigidity of the isotropic plate, so that equation of motion (\ref{eq:untransformed-plate-eq}) simplifies to
\begin{equation}
D^{(0)} \GD^2 w - \rho h \omega^2 w = 0.
\label{eq:IH-plate}
\end{equation}

Consider the radial invertible ``push-out'' transformation, introduced in \cite{pendry2006,schurig2006,leonhardt2006,GreLasUhl2003}.
Within  $R_1<r<R_2$, the transformation $\vect{x} = \mathcal{F}(\vec{X})$ is given by
\begin{equation}
            r=R_1+\dfrac{(R_2-R_1)}{R_2}  R,
           \,\,\, \theta=\Theta,
  \quad \text{when} \quad 0\leq R \leq R_2,
\label{eq:circulartrans}
\end{equation}
where $\vec{X}=(R,\Theta)^T$ and $\vect{x}=(r,\theta)^T$.

The Jacobi matrix ${\bf F}$ in cylindrical coordinates $(r,\theta,z)$ has the form
\begin{equation}
{\bf F}= 
\frac{R_2-R_1}{R_1}{\bf e}_r\otimes{\bf e}_r+\frac{R_2-R_1}{R_1}\frac{r}{r-R_1}{\bf e}_\theta\otimes{\bf e}_\theta,
\label{eq:Jacobi}
\end{equation}
where ${\bf e}_r={\bf e}_R$, ${\bf e}_\theta={\bf e}_\Theta$
is the orthonormal basis and $\otimes$ stands for the dyadic product.

By direct application of the transformation or alternatively a double application of~\cite[Lemma 2.1]{norris2008} the isotropic equation, in new polar coordinates, may be expressed as
\begin{multline}
\frac{(r-R_1)^2}{r^2}\frac{\partial^4 w}{\partial r^4}+\frac{2}{r^2}\frac{\partial^4 w}{\partial r^2 \partial \theta^2}+\frac{1}{r^2(r-R_1)^2}\frac{\partial^4 w}{\partial \theta^4}+\frac{2(r-R_1)}{r^2}\frac{\partial^3 w}{\partial r^3}-\frac{2}{r^2(r-R_1)}\frac{\partial^3 w}{\partial r \partial \theta^2} \\ -\frac{1}{r^2}\frac{\partial^2 w}{\partial r^2 }+\frac{4}{r^2(r-R_1)^2}\frac{\partial^2 w}{ \partial \theta^2}+\frac{1}{r^2(r-R_1)}\frac{\partial w}{\partial r}- \frac{\rho R_2^4(r-R_1)^2}{D^{(0)}r^2(R_2-R_1)^4}h \omega^2w=0.
\label{eq:transformed1}
\end{multline}

Letting 
\beq
D'_r=D^{(0)}\frac{(r-R_1)^2}{r^2}, \,\,\, D'_{r\theta}=D^{(0)} ~ \mbox{and} ~ D'_\theta=D^{(0)}\frac{r^2}{(r-R_1)^2},  
\eequ{rigidities}
equation \eq{eq:transformed1} may be re-written as

\begin{multline}
D'_r\frac{\partial^4 w}{\partial r^4}+\frac{2}{r^2}D'_{r \theta}\frac{\partial^4 w}{\partial r^2 \partial \theta^2}+\frac{1}{r^4}D'_\theta\frac{\partial^4 w}{\partial \theta^4}+\frac{2}{r}D'_r\frac{r}{(r-R_1)}\frac{\partial^3 w}{\partial r^3}-\frac{2}{r^3}D'_{r\theta}\frac{r}{(r-R_1)}\frac{\partial^3 w}{\partial r \partial \theta^2} \\ -\frac{1}{r^2}D'_\theta\frac{(r-R_1)^2}{r^2}\frac{\partial^2 w}{\partial r^2 }+\frac{2}{r^4}(D'_\theta+D'_{r\theta})\frac{2r^2}{r^2+(r-R_1)^2}\frac{\partial^2 w}{ \partial \theta^2}\\+\frac{1}{r^3}D'_\theta\frac{(r-R_1)}{r}\frac{\partial w}{\partial r}- \frac{\rho R_2^4(r-R_1)^2}{r^2(R_2-R_1)^4}h \omega^2w=0.
\label{eq:transformed2}
\end{multline}

If we introduce the notation $\rho'=\rho R_2^4/(R_2-R_1)^4$ for the normalised mass density, then it is tempting to assume that \eq{eq:transformed2} represents an orthotropic inhomogenous plate with the stiffness rigidities \eq{rigidities}. The question is: can such an assumption be justified?

On one hand, the fourth-order terms in \eq{eq:transformed2} agree with the structure of \eq{eq:untransformed-plate-eq}. On the other hand, the additional lower-order terms have to be analysed.

We also note that, after the normalisation, equation (\ref{eq:transformed1}) can be written in the compact form
\begin{equation}
\hat\nabla^2_{R_1}(\hat\nabla^2_{R_1}w) - \frac{\rho R_2^4}{D^{(0)}(R_2-R_1)^4}h \omega^2w=0,
\label{eq030}
\end{equation}
where the differential operator $\hat\nabla^2_{R_1}$ is defined as
\begin{equation}
\hat\nabla^2_{R_1}=\frac{1}{r-R_1}\frac{\partial}{\partial r}\left[(r-R_1)\frac{\partial}{\partial r}\right]+\frac{1}{(r-R_1)^2}\frac{\partial^2}{\partial \theta^2}.
\label{eq030a}
\end{equation}
We would like to emphasise that for $R_1 >0$, the operators $\hat\nabla^2_{R_1}$  and $\nabla^2$ are not the same. The operator $\hat\nabla^2_{R_1}$ will be referred to as the `shifted Laplace operator', which becomes the classical Laplace's operator only when $R_1=0$, i.e. in the absence of the cloak. Correspondingly, we will use the terms `shifted Helmholtz' and `shifted modified Helmholtz' for the operators $\hat\nabla^2_{R_1} + \beta^2$ and $\hat\nabla^2_{R_1} - \beta^2$, respectively.
 
Following the representation (\ref{eq030}) we can express the transformed equation in the form
\begin{equation}
\hat\nabla^2_{R_1}(\hat\nabla^2_{R_1}w) - \beta^4 w=(\hat\nabla^2_{R_1} + \beta^2)(\hat\nabla^2_{R_1} - \beta^2) w= 0,
\label{eq031a}
\end{equation}
where 
\begin{equation}
\beta=\frac{R_2}{R_2-R_1}\left(\frac{\rho h}{D^{(0)}}\omega^2\right)^{1/4}
\label{eq031b}
\end{equation}
has the physical dimension of $[\mbox{m}^{-1}]$.
The solution of equation (\ref{eq031a}) admits the representation
\begin{equation}
w(r,\theta)=w_{HS}(r,\theta)+w_{MS}(r,\theta),
\label{eq032a}
\end{equation}
where
\begin{equation}
(\hat\nabla^2_{R_1} + \beta^2)w_{HS}(r,\theta)=0, \quad
(\hat\nabla^2_{R_1} - \beta^2)w_{MS}(r,\theta)=0, \quad
\label{eq033a}
\end{equation}
and hence $w$ is the superposition of waves of shifted Helmholtz type $w_{HS}$ and shifted modified Helmholtz type $w_{MS}$.
A semi-analytical solution can be found by implementing the series representation
\begin{equation}
w(r,\theta)=\sum_{n=0}^{+\infty} w_n(r) e^{in\theta},
\label{eq034a}
\end{equation}
where
\begin{equation}
w_n(r)=A_n J_n[\beta(r-R_1)]+E_nH_n^{(1)}[\beta(r-R_1)]+B_nI_n[\beta(r-R_1)]+F_nK_n[\beta(r-R_1)].
\label{eq035a}
\end{equation}
In equation (\ref{eq035a}) $J_n$ is the Bessel function, $H_n^{(1)}$is the Hankel function, and $I_n$ and
$K_n$ are the modified Bessel functions related to $J_n$ and $H_n^{(1)}$ by
\begin{equation}
I_n(z)=i^{-n} J_n(iz), \quad
K_n(z)=\frac{\pi i^{n+1}}{2}H_n^{(1)}(iz),
\label{eq036a}
\end{equation}
respectively (see \cite{AbramowitzStegun1965}, equations 9.6.3 and 9.6.4). The coefficients of the expansion (\ref{eq034a}), (\ref{eq035a}) are determined from the boundary and the interface conditions on the contour of the cloak.

\section{Transformation cloaking for a membrane versus flexural plate}

 The radial ``push-out'' transformation \eq{eq:circulartrans} can be used
 to design a cloak that will route an incident wave around a finite-size obstacle in an elastic membrane. Norris \cite{norris2008} has discussed this problem in detail. The governing equation for a time-harmonic out-of-plane displacement $u$ of an elastic membrane has the form
\beq
\left(\nabla_{\vec{X}}\cdot\mu\nabla_{\vec{X}} + \rho\omega^2\right)u(\vec{X}) = 0,\; \vec{X}\in\mathbb{R}^2,
\eequ{membr1}
where $\mu$ stands for the stiffness matrix and $\rho$ and $\omega$ are the mass density and the radian frequency, respectively.
If an invertible mapping $\vect{x} = \CF(\vec{X})$ is applied within the cloaking region, then the transformed equation becomes
\beq
\left(\nabla\cdot\mu\vec{C}(\vect{x})\nabla + \frac{\rho\omega^2}{J(\vec{x})}\right)u(\vec{x}) = 0,
\eequ{membr2}
where
\beq
\vec{C} = \frac{\vec{F}\vec{F}^\mathrm{T}}{J},\quad {\bf F} = \nabla_{\vec{X}}{\vect{x}},\quad J = \det \vec{F}.
\eequ{membr3}
It is important to note that equation \eq{membr2}, similar to \eq{membr1},  describes a vibrating membrane, but with different elastic stiffness and a non-uniform distribution of mass across the transformed region.

In contrast, for the model of a flexural plate, 
equation \eq{eq:transformed2}, after the transformation 
\eq{eq:circulartrans}, 
does not preserve the physical interpretation, i.e. it is no longer the equation of free vibrations of a plate.
This suggests that the problem in hand is very different from the model of a cloak for a membrane. It presents an additional challenge to identify the physical configuration consistent with the new equation \eq{eq:transformed2}. This issue is to be discussed in the next section.

\section{The cloaking transformation does not produce an orthotropic inhomogeneous  plate}

We make a direct comparison between the transformed equation \eqref{eq:transformed2}, and the equation for an inhomogeneous orthotropic plate. Firstly, we note that the moments $M_r, M_\theta$ and $M_{\theta r}$ satisfy the partial differential equation:
\begin{equation}
\fr{\prt^2 M_r}{\prt r^2} + \fr{2}{r} \fr{\prt M_r}{\prt r} + \fr{2}{r} \fr{\prt^2 M_{\theta r}}{\prt r \prt \theta} +
\fr{2}{r^2} \fr{\prt M_{\theta r}}{\prt \theta} + \fr{1}{r^2} \fr{\prt^2 M_{\theta}}{\prt \theta^2} - \fr{1}{r} \fr{\prt M_\theta}{\prt r} - \rho h \fr{\prt^2 w}{\prt t^2} =0.
\label{eq020}
\end{equation}

For an orthotropic inhomogeneous plate, where rigidities and Poisson's coefficients vary radially, equation (\ref{eq020}) has the form
\begin{eqnarray}
\nonumber
D_r\left( \frac{\partial^4w}{\partial r^4}+\frac{2}{r}\frac{\partial^3w}{\partial r^3}\right)+
\frac{2D_{r\theta}}{r^2}\left( \frac{\partial^4w}{\partial r^2\partial \theta^2}-\frac{1}{r}\frac{\partial^3w}{\partial r\partial \theta^2}+\frac{1}{r^2}\frac{\partial^2w}{\partial \theta^2}\right)\\
\nonumber
+\frac{D_{\theta}}{r^2}\left( \frac{1}{r^2}\frac{\partial^4w}{\partial \theta^4}+\frac{2}{r^2}\frac{\partial^2w}{\partial \theta^2}
-\frac{\partial^2w}{\partial r^2}+\frac{1}{r}\frac{\partial w}{\partial r}\right)+
2\frac{\partial D_r}{\partial r}\left(\frac{\partial^3w}{\partial r^3}+\frac{1}{r}\frac{\partial^2w}{\partial r^2}\right)\\
\nonumber
+\frac{2}{r^2}\frac{\partial D_{r\theta}}{\partial r}\left(\frac{\partial^3w}{\partial r\partial \theta^2}-\frac{1}{r}\frac{\partial^2w}{\partial \theta^2}\right)+
\frac{1}{r}\frac{\partial (D_r\nu_\theta)}{\partial r}\frac{\partial^2w}{\partial r^2}-
\frac{1}{r^2}\frac{\partial D_\theta}{\partial r}\left(\frac{\partial w}{\partial r}+\frac{1}{r}\frac{\partial^2w}{\partial \theta^2}\right)\\
+\frac{\partial^2 D_r}{\partial r^2}\frac{\partial^2w}{\partial r^2}+
\frac{1}{r}\frac{\partial^2 (D_r\nu_\theta)}{\partial r^2}\left(\frac{\partial w}{\partial r}+\frac{1}{r}\frac{\partial^2w}{\partial \theta^2}\right)+ \rho h \fr{\prt^2 w}{\prt t^2} =0.
\label{eq021}
\end{eqnarray}
Direct comparison of \eq{eq021} and \eq{eq:transformed2} shows that the fourth order terms agree in the equation of the orthotropic plate and the transformed equation within the cloaking region. However, a discrepancy occurs in other lower order terms, and hence the transformed equation \eqref{eq:transformed2} does not represent a classical orthotropic Kirchhoff plate. Additional physical constraints are needed to complete the model. This will be achieved through an approximation discussed in the next section. 

\section{The cloaking approximation}

Equation \eq{eq021} can be rewritten after the substitution of flexural rigidity coefficients as in \eq{rigidities}:
\begin{eqnarray}
\nonumber
\frac{(r-R_1)^2}{r^2}\frac{\partial^4 w }{\partial r^4}+
\frac{2}{r^2}\frac{\partial^4 w }{\partial r^2\partial \theta^2}+
\frac{1}{(r-R_1)^2r^2}\frac{\partial^4 w }{\partial \theta^4}+
\frac{2(r^2-R_1^2)}{r^3}\frac{\partial^3 w }{\partial r^3}-
\frac{2}{r^3}\frac{\partial^3 w }{\partial r\partial \theta^2}\\
\nonumber
+\frac{r^4R_1+2r^3R_1^2-6r^2R_1^3+6rR_1^4-r^5-2R_1^5-2\nu_r r^4 R_1}{(r-R_1)^3r^4}\frac{\partial^2 w }{\partial r^2}\\
\nonumber
+2\frac{(r-R_1)(2r-R_1)(r^2-r R_1+R_1^2)+\nu_r r^2R_1(2r+R_1)}{(r-R_1)^4 r^4}\frac{\partial^2 w }{\partial \theta^2}\\
+\frac{r^2-R_1^2+2\nu_r R_1(2r+R_1)}{r(r-R_1)^4}\frac{\partial w }{\partial r}-
\rho h\omega^2\frac{r_2^4(r-R_1)^2}{r^2(R_2-R_1)^4}w=0.
\label{eq022}
\end{eqnarray}
Direct comparison with equation \eq{eq:transformed1} shows the discrepancy in the third-order derivative terms in addition to that in the lower-order terms.
The difference between the left-hand sides of equations \eq{eq:transformed1} and \eq{eq022} is
\begin{eqnarray}
\nonumber
\frac{2R_1(r-R_1)}{r^3}\frac{\partial^3 w }{\partial r^3}+\frac{2R_1}{r^3(r-R_1)}\frac{\partial^3 w }{\partial r\partial \theta^2}\\
\nonumber
-\frac{R_1\left(2R_1^4+2r^4-5r^3R_1+7r^2R_1^2-6rR_1^3+2\nu_r r^4\right)}{r^4(r-R_1)^3}\frac{\partial^2 w }{\partial r^2}\\
\nonumber
+\frac{2R_1\left(2\nu_rr^3+\nu_rr^2R_1-r^3+4r^2R_1-4rR_1^2+R_1^3\right)}{(r-R_1)^4r^4}\frac{\partial^2 w }{\partial \theta^2}\\
+\frac{R_1\left(4\nu_r r^2+2\nu_rrR_1-4r R_1+3r^2+R_1^2\right)}{r^2(r-R_1)^4}\frac{\partial w }{\partial r} .
\label{eq023}
\end{eqnarray}
It is apparent that all coefficients in the above equation have the form $f_j(r) R_1/r$, with $f_j$ being smooth functions when $r> R_1$, and these coefficients are small when $R_1/r$ is considered as a small parameter, in particular, when the penetration depth for the incident wave into the cloaking region is small.

In the approximation implemented here, we chose the parameters of the cloak in such a way that the interior diameter of the cloaking ring is sufficiently small compared to the diameter of the whole cloaking region and compared to the wavelength of the incident wave, i.e. the following non-dimensional quantities are small
$$
R_1 / R_2 \ll 1, \quad \Gb R_1 \ll 1,
$$
where $\beta$ is defined by \eq{eq031b}.
The material outside the cloak remains unaffected by the transformation, whereas the interior material represents a radially orthotropic plate in the framework of the approximation described here (see equations \eq{eq022}, \eq{eq023}).  
Numerical simulations below show the efficiency of our concept, which is also in  agreement with the experimental evidence published in \cite{stenger2012}.

\subsection{Numerical illustration}
\label{sec:numerics}

The notion of an approximate cloak, introduced above, is used here in the numerical illustrations. This approximation is valid for a certain choice of geometrical parameters and frequency values.

Numerical simulations are produced for an elastic, isotropic Kirchhoff plate which contains a radially orthotropic inhomogeneous cloaking layer.
Without loss of generality, the incident field is represented by a flexural plane wave propagating horizontally. Perfectly matched layers (PML) are used on the 
exterior boundary of the computational domain.
PML conditions are ``absorbing'' boundary conditions simulating a non-reflective exterior contour.
The
parameters used for the numerical simulations are shown in table~\ref{tab:parameters}.
The exterior of the cloak corresponds to a homogeneous isotopic plate, whereas the interior of the cloak is an inhomogeneous radially orthotropic plate.
The numerical simulations were produced using Comsol Multiphysics\textsuperscript{\textregistered} (see Appendix \ref{App1} for more details on the numerical implementation).

\begin{table}[htb]
\centering
\begin{tabular}{c@{\qquad}c@{\qquad}c@{\qquad}}
\toprule
Parameter & \multicolumn{2}{c}{Value} \\
		 & Exterior of the cloak & Interior of the cloak \\
		 \midrule
$D^{(0)}$		 & 1		&  1 \\
$D_r$		 & 1		&  $\tfrac{(r-R_1)^2}{r^2}$ \\
$D_\theta$	 & 1		&  $\tfrac{r^2}{(r-R_1)^2}$ \\
$D_{r\theta}$	 & 1		&  1 \\
$\nu_r$		 & 0.3	& 0.3 \\
$\nu_\theta$	 & 0.3	& $\tfrac{3r^4}{10(r-R_1)^4}$ \\[1.2 mm]
$\rho$		 & 1	 	& $\tfrac{R_2^4(r-R_1)^2}{r^2(R_2-R_1)^4}$\\
$h$                      & $0.001$ & $0.001$\\ 
\bottomrule
\end{tabular}
\caption{\label{tab:parameters}
The parameters used in the numerical simulations, see \eq{rigidities}.}
\end{table}

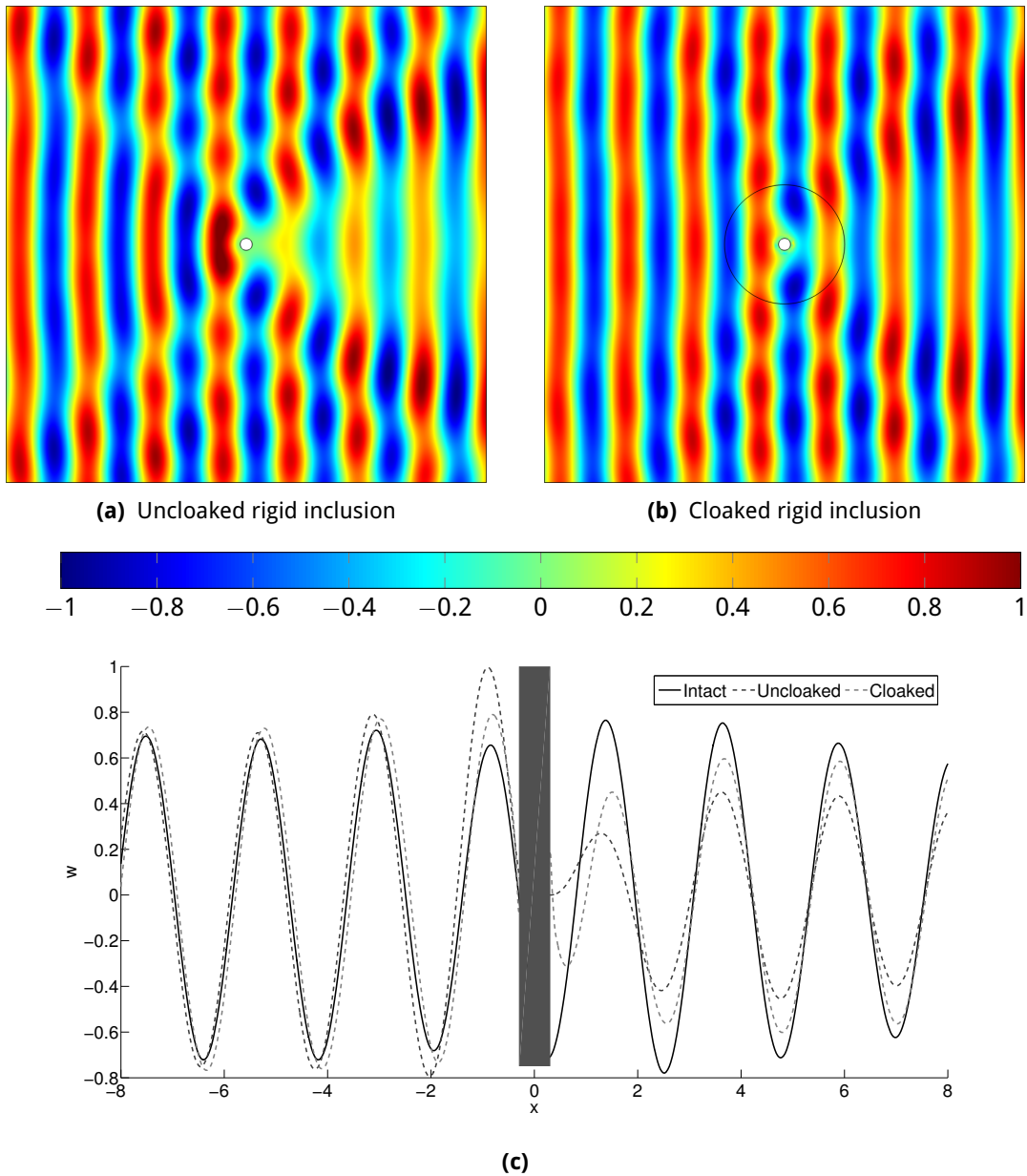
\begin{figure}
\centering
\begin{subfigure}[c]{0.45\textwidth}
\includegraphics[width=\linewidth]{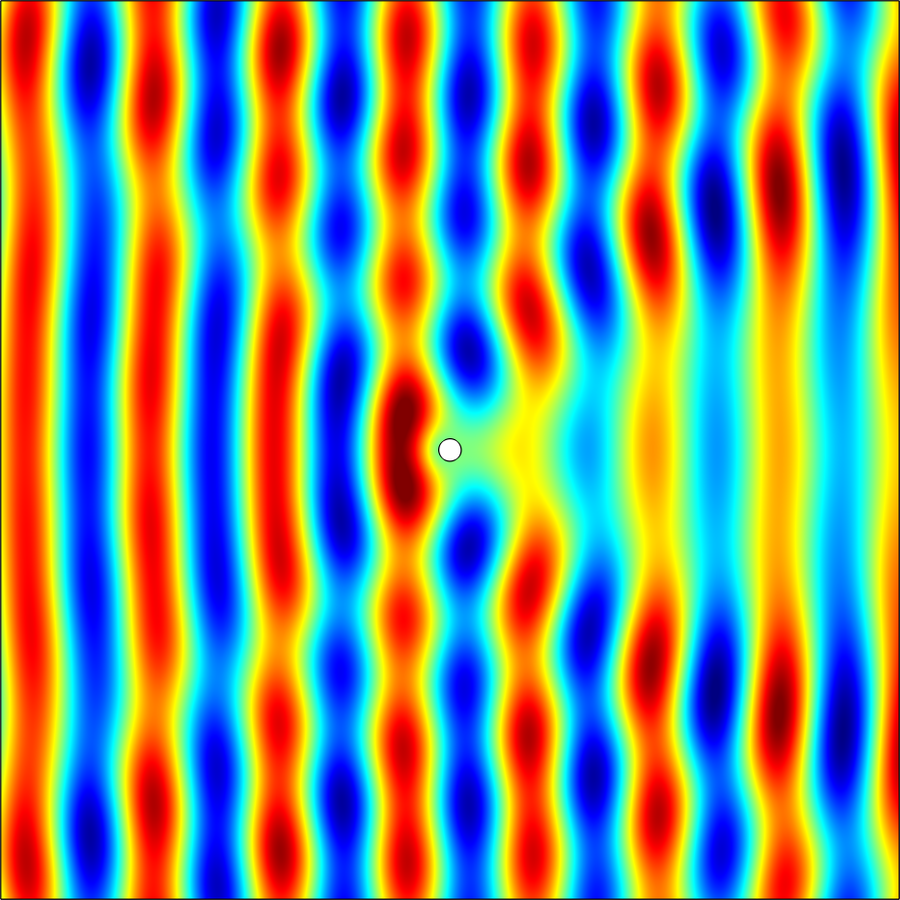}
\caption{\label{fig:Uncloaked-40-0p2}
Uncloaked rigid inclusion}
\end{subfigure}
\qquad
\begin{subfigure}[c]{0.45\textwidth}
\includegraphics[width=\linewidth]{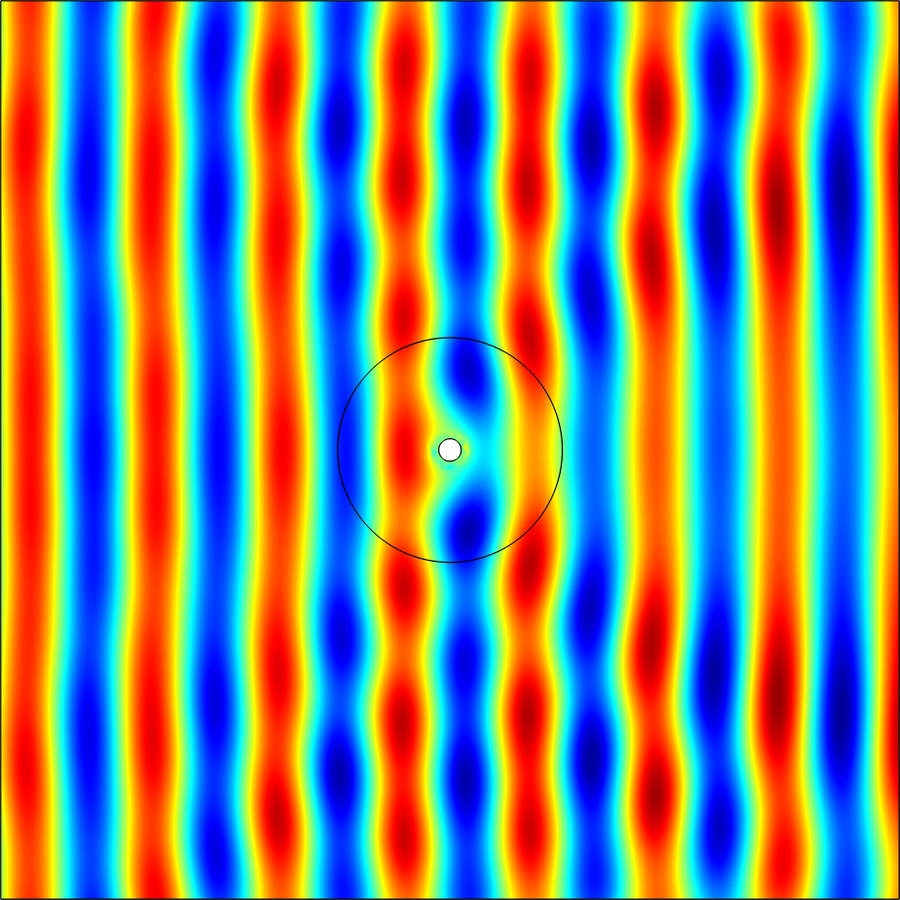}
\caption{\label{fig:Cloaked-40-0p2}
Cloaked rigid inclusion}
\end{subfigure}
\begin{subfigure}[c]{0.9\textwidth}
\centering
\begin{tikzpicture}
\begin{axis}[
    hide axis,
    scale only axis,
    height=0pt,
    width=0pt,
    colormap/jet,
    colorbar horizontal,
    point meta min=-1,
    point meta max=1,
    colorbar style={
        width=\linewidth,
    }]
    \addplot [draw=none] coordinates {(0,0)};
\end{axis}
\end{tikzpicture}
\end{subfigure}
\qquad
\begin{subfigure}[c]{\textwidth}
\includegraphics[width=\linewidth]{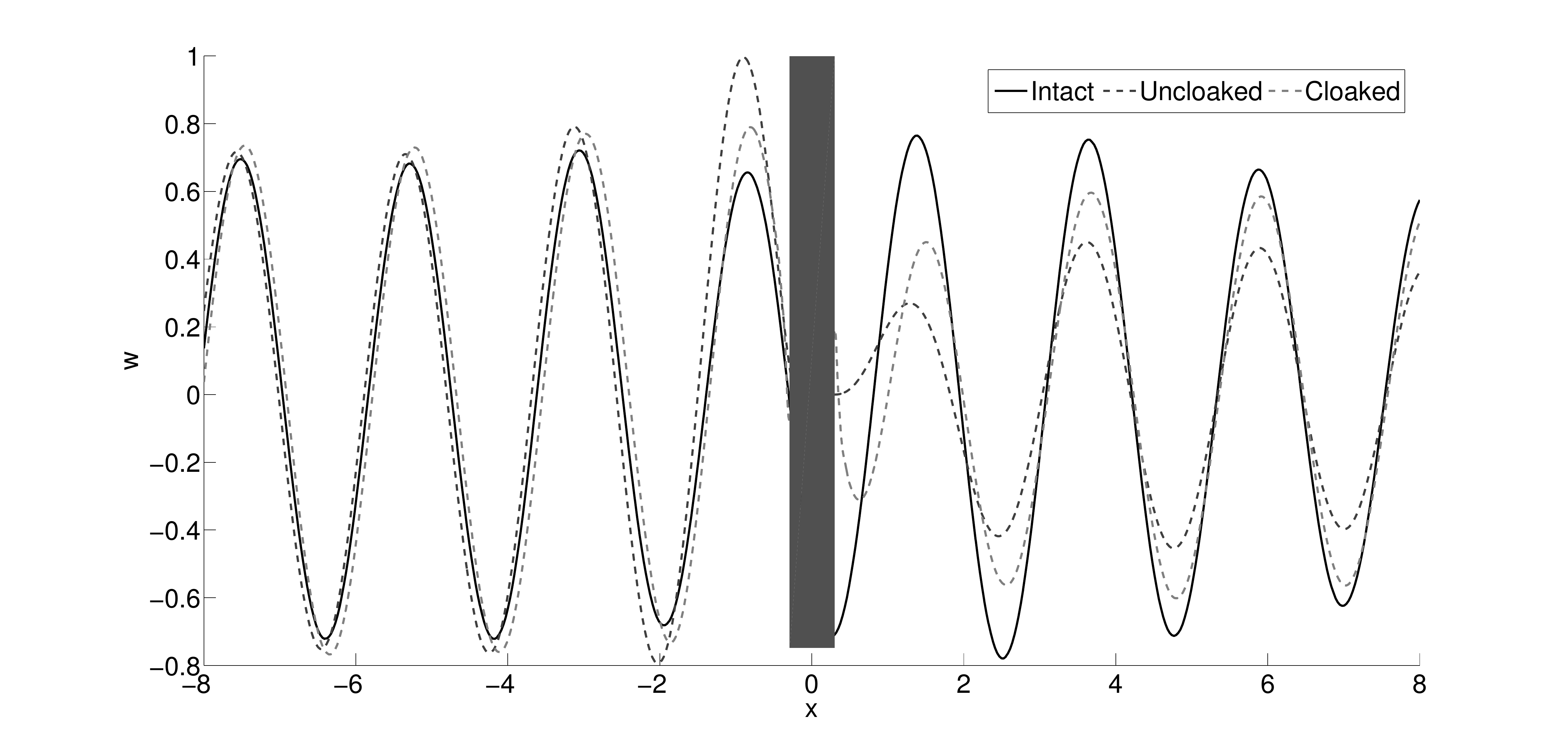}
\caption{\label{fig:Comp-40-0p2}}
\end{subfigure}
\caption{\label{fig:Cloaking-40-0p2}
The flexural displacement $w(\vect{x})$ generated by a line source in the far field.
Parts (a) and (b) show the field for an uncloaked and cloaked rigid inclusion respectively.
Part (c) shows the flexural displacement for cases (a) and (b) together with the flexural displacement in the absence of both inclusion and cloak along a line passing through the centre of the inclusion in the direction of the incident wave.
The rigid inclusion is indicated by the grey rectangle in part (c).
The non-dimensional radian frequency $\omega=40$ and the radii of the cloak and inclusion are $R_2 = 2$ and $R_1=0.2$, respectively.}
\end{figure}

In figure  \ref{fig:Cloaking-40-0p2}, we consider the case of  interior and exterior radii for the cloaking region chosen as $R_1=0.2$ and $R_2 = 2 $.
The normalised radian frequency is  $\Go=40$. 
Part (a) of  figure  \ref{fig:Cloaking-40-0p2} shows the uncloaked inclusion, and part (b) of the same figure shows the cloaked coated inclusion, where the shadow region has been significantly suppressed. 
Part (c) of figure~\ref{fig:Cloaking-40-0p2} shows the flexural displacement, for cases (a) and (b) together with the field in the absence of both cloak and inclusion;
here the field is plotted along a line passing through the centre of the inclusion in the direction of the incident wave.
For this choice of parameters,
we observe good cloaking of a finite object for the incident plane flexural wave.

\begin{figure}
\centering
\begin{subfigure}[c]{0.45\textwidth}
\includegraphics[width=\linewidth]{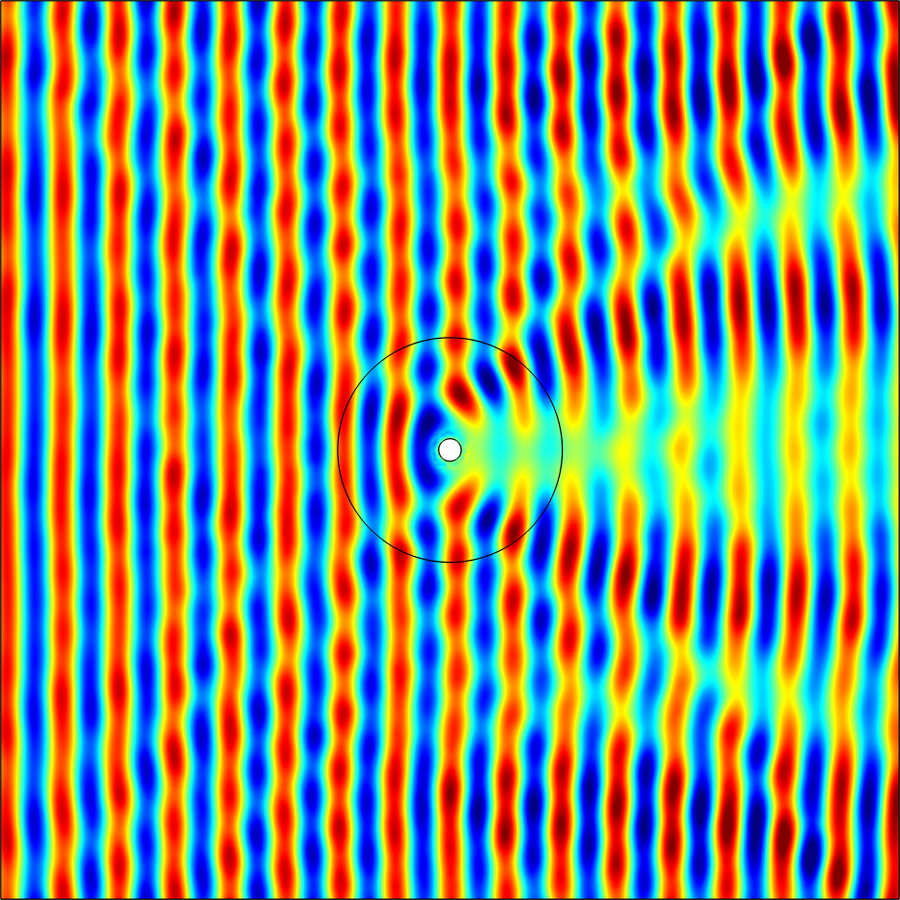}
\caption{\label{fig:Cloaked-40-0p4}
Higher frequency regime}
\end{subfigure}
\qquad
\begin{subfigure}[c]{0.45\textwidth}
\includegraphics[width=\linewidth]{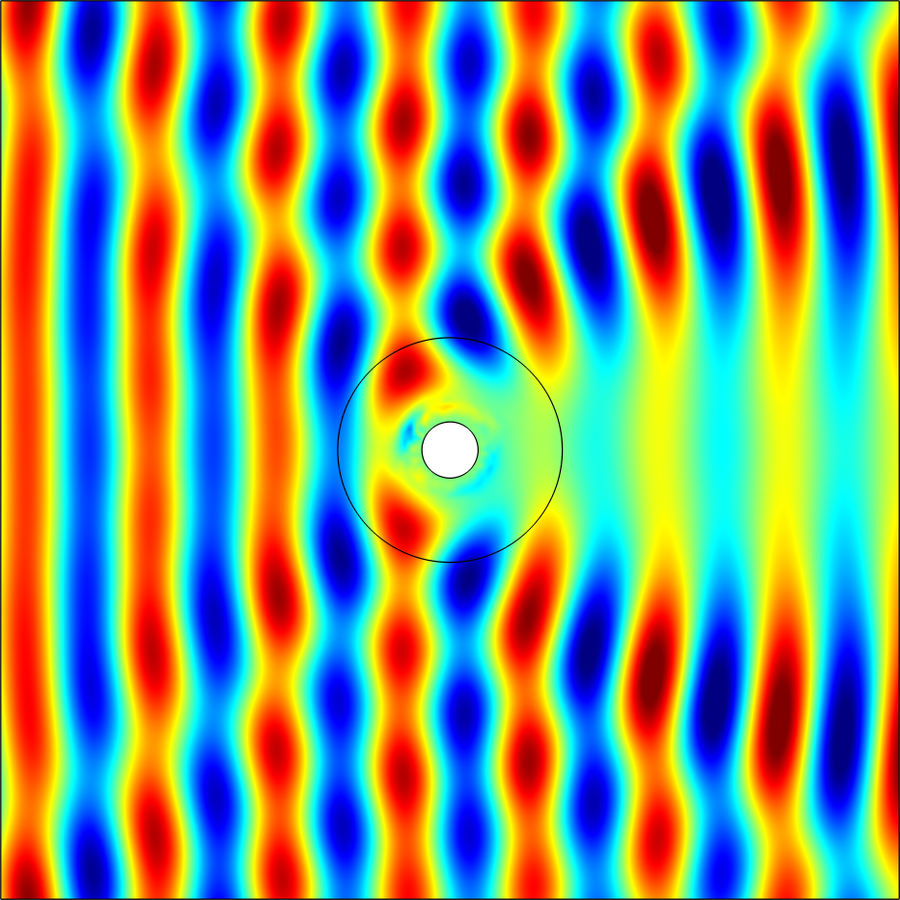}
\caption{\label{fig:Cloaked-200-0p2}
Larger $R_1/R_2$ ratio}
\end{subfigure}
\caption{\label{fig:Cloaking-200-0p2}
The flexural displacement $w(\vect{x})$ generated by a line source subject to the scattering by the coated  rigid inclusion.
In the cases shown the coating does not provide cloaking.
In part (a), the incident wave has a larger frequency than in figure \ref{fig:Cloaking-40-0p2}: the non-dimensional radian frequency is $\omega=200$.
In part (b), the frequency of the incident wave is the same as in figure \ref{fig:Cloaking-40-0p2} ($\omega=40$), but the ratio of the interior and exterior radii of the coating has doubled: in the present case $R_2 = 2$ and $R_1=0.4.$
Cloaking has been affected in these two cases.
The colour range is as indicated in figure~\ref{fig:Cloaking-40-0p2}.}
\end{figure}

It is also expected that the approximation is frequency sensitive, and the properties of the approximate cloak may also change with the variation of the thickness of the cloak. 
This is illustrated in figure  \ref{fig:Cloaking-200-0p2}.  In part (a) of that figure, the simulation corresponds to the case of a higher frequency ($\omega=200$), and the cloaked obstacle shows a non-suppressed shadow. Similarly,  in part (b) of figure \ref{fig:Cloaking-200-0p2} we have  non-suppressed shadow for a different reason. 
Although the frequency of the incident wave remains the same as in figure \ref{fig:Cloaking-40-0p2}, the size of the obstacle has increased and the interior radius of the cloak is twice as large as the case in figure \ref{fig:Cloaking-40-0p2}(b). Consequently, in both diagrams shown in figure \ref{fig:Cloaking-200-0p2}
the cloaking has been affected.

\section{Alternative approach: plate subjected to in-plane forces and pre-stress}
\label{prestress_section}

In this section we show that, by choosing a different normalisation, it is possible to give a physical interpretation of the transformed plate equations as  a Kirchhoff plate subjected to in-plane forces in addition to the usual flexural behavior. This can also lead to a broadband perfect cloak. Here, we extend the cartesian formulation given recently in Colquitt et al. \cite{ColBruGeiMovMovJon2014} to the cylindrical cloak configuration.
In particular, equations (\ref{eq:transformed1}) and (\ref{eq030}) are normalised in the following way
\begin{eqnarray}
\nonumber
D^{(0)}\frac{r-R_1}{r}\hat\nabla^2_{R_1}(\hat\nabla^2_{R_1}w) - \frac{\rho R_2^4(r-R_1)}{r(R_2-R_1)^4}h \omega^2w=\\
\nonumber
D^{(0)}\left[\frac{r-R_1}{r}\frac{\partial^4 w}{\partial r^4}+\frac{2}{r(r-R_1)}\frac{\partial^4 w}{\partial r^2 \partial \theta^2}+\frac{1}{r(r-R_1)^3}\frac{\partial^4 w}{\partial \theta^4}+\frac{2}{r}\frac{\partial^3 w}{\partial r^3}-\frac{2}{r(r-R_1)^2}\frac{\partial^3 w}{\partial r \partial \theta^2} \right. \\ 
\left.
-\frac{1}{r(r-R_1)}\frac{\partial^2 w}{\partial r^2 }+\frac{4}{r(r-R_1)^3}\frac{\partial^2 w}{ \partial \theta^2}+\frac{1}{r(r-R_1)^2}\frac{\partial w}{\partial r}\right]- \frac{\rho R_2^4(r-R_1)}{r(R_2-R_1)^4}h \omega^2w=0.
\label{eq200}
\end{eqnarray}

Then, introducing the following definition for the rigidity and inertial parameters
\begin{eqnarray}
\nonumber
D_r=D^{(0)}\frac{r-R_1}{r},\quad
D_\theta=D^{(0)}\left(\frac{r}{r-R_1}\right)^3,\quad
D_{r\theta}=D^{(0)}\frac{r}{r-R_1}, \\ 
\nu_{\theta}=\nu_{r}^{-1}=\left(\frac{r}{r-R_1}\right)^2, \quad
\rho'=\frac{\rho R_2^4(r-R_1)}{r(R_2-R_1)^4},
\label{eq201}
\end{eqnarray}
the equations (\ref{eq020}) and (\ref{eq021}), for an inhomogeneous orthotropic plate, match all the terms involving fourth-, third- and zero-order derivatives of the transverse displacements in equation (\ref{eq200}).
The remaining terms (second- and first-order) can finally be matched by considering additional pre-stress forces ${\bf N}$ and in-plane body forces ${\bf S}$, having components
\begin{eqnarray}
\nonumber
N_{rr}=\frac{3r-2R_1}{r(r-R_1)^3}R_1,\quad
N_{\theta\theta}=-\frac{3rR_1}{(r-R_1)^4},\quad
N_{r\theta}=0,\\
S_r=\frac{3R_1}{r(r-R_1)^3},\quad S_\theta=0.
\label{eq202}
\end{eqnarray}
These are constrained to satisfy the in-plane balance equation
\begin{equation}
\nabla\cdot{\bf N}+{\bf S}=0.
\label{eq203}
\end{equation}

The final form for the transformed equation is
\begin{equation}
\nabla\cdot(\nabla\cdot {\bf M})+{\bf N}:\nabla\nabla w-{\bf S}\cdot\nabla w=-\rho'h\omega^2 w.
\end{equation}
leading to a consistent physical interpretation.
In Colquitt et al. \cite{ColBruGeiMovMovJon2014}, for a different cloak geometry, we have shown that such a pre-stressed elastic system leads to broadband cloaking.

We would like to note the resemblance of the above computations  in figures \ref{fig:Cloaking-40-0p2} and \ref{fig:Cloaking-200-0p2} with those in paper \cite{stenger2012}, which shows results from an experimental study of a structured cloak and flexural waves. In \cite{stenger2012} 
the cloaking approximation is shown to be frequency sensitive, so that cloaking does not occur for frequencies above a certain threshold. 
From  \eq{eq:untransformed-plate-eq}--\eq{eq:untransformed-plate-eq-a}, it is clear that four independent elastic parameters $D_r$, $D_\theta$, $D_{r\theta}$, and $\nu_r$ are required to characterise a radially orthotropic plate (also see the classical papers \cite{Carrier1943,Carrier1944}).
Only Young's moduli $E_r$ and $E_\theta$ appear to be given in paper \cite{stenger2012}.

The inertial properties are defined by the mass density which is also required for the computations and experiment. Different normalisation of the mass density can be applied; in particular, the mass density used in \cite{farhat2009PRL,stenger2012} was constant.
Here we have defined all of the required parameters and explained how they fit into the configuration approximating the flexural cloak. We have also given the range of validity of 
such an approximation.

\section{Asymptotic derivation of the transformed plate equation from the equations of elasticity}
\label{Sect03}

In this section, the transformed equations of motion for the Kirchhoff plate (\ref{eq030}) are deduced directly from the transformed equations of motion of three-dimensional linear elasticity. An asymptotic model is implemented in order to obtain the lower-dimensional plate model from the analysis of a thin three-dimensional solid. 
It was shown in earlier works \cite{norris2011,milton2006,brun2009}  that the transformed equations of elasticity are subject to the choice of gauge. In particular, the resulting material may lack the minor symmetries in the constitutive equations. This does not occur in the case of flexural plates, as demonstrated below.  

\subsection{Transformed equation of elasticity}
The Navier equations 
\begin{equation}
(\lambda+2\mu)\nabla_{\vec{X}}(\nabla_{\vec{X}}\cdot{\bf U})-\mu\nabla_{\vec{X}}\times(\nabla_{\vec{X}}\times{\bf U})=\rho\frac{\partial^2 {\bf U}}{\partial t^2}
\label{eqn000}
\end{equation}
describing the displacement field 
${\bf U}={\bf U}({\vec{X}},t)=\left(U_R,\, U_\Theta ,\, U_Z \right)^T$, with ${\bf X}=(R,\,\Theta,\,Z)^T$, in a linear elastic and isotropic medium can be conveniently expressed in cylindrical coordinates, i.e.
\begin{eqnarray}
\label{eqn001}
\nonumber
\left(\lambda+\mu\right)\frac{\partial}{\partial R}(\nabla_\vec{X}\cdot{\bf U})+\mu\left(\nabla^2_\vec{X}U_R-\frac{U_R}{R^2}-\frac{2}{R^2}\frac{\partial U_\Theta}{\partial \Theta}\right)=\rho\frac{\partial^2 U_R}{\partial t^2},& \\
\nonumber
\frac{\lambda+\mu}{R}\frac{\partial}{\partial \Theta}(\nabla_\vec{X}\cdot{\bf U})+\mu\left(\nabla^2_\vec{X}U_\Theta-\frac{U_\Theta}{R^2}+\frac{2}{R^2}\frac{\partial U_R}{\partial \Theta}\right)=\rho\frac{\partial^2 U_\Theta}{\partial t^2},&\quad \vec{X}\in\chi,\\
\left(\lambda+\mu\right)\frac{\partial}{\partial Z}(\nabla_\vec{X}\cdot{\bf U})+\mu\nabla^2_\vec{X}U_Z=\rho\frac{\partial^2 U_Z}{\partial t^2}.&
\end{eqnarray}
In \eqref{eqn001}, $\chi=\Omega \times [-h/2, h/2]$, with $\Omega\subseteq\mathbb{R}^2$, $\lambda$ and $\mu$ are the Lam\'{e} moduli, $\rho$ is the mass density
of the medium and zero body forces are assumed. . 
Field equations \eqref{eqn001} are accompanied by homogeneous Neumann boundary conditions on the upper and lower external surfaces $Z=\pm h/2$:
\begin{eqnarray}
\label{eqn002}
\nonumber
\mu\left(\frac{\partial U_R}{\partial Z}+\frac{\partial U_Z}{\partial R}\right)=0,\\
\nonumber
\mu\left(\frac{1}{R}\frac{\partial U_Z}{\partial \Theta}+\frac{\partial U_\Theta}{\partial Z}\right)=0,\\
\lambda\left(\frac{\partial U_R}{\partial R}+\frac{1}{R}\frac{\partial U_\Theta}{\partial \Theta}+\frac{U_R}{R}\right)+(\lambda+2\mu)\frac{\partial U_Z}{\partial Z}=0,
\end{eqnarray}
where $(R,\Theta)\in \Omega$.

Now, we introduce a geometric transformation 
\begin{equation}
            r=R_1+\dfrac{(R_2-R_1)}{R_2}  R,
           \,\,\, \theta=\Theta,
           \,\,\, z=Z 
  \quad \text{when} \quad 0\leq R \leq R_2.
\label{eqn002a}
\end{equation}
Accordingly, 
the Jacobi matrix ${\bf F}$ (in cylindrical coordinates) and the Jacobian $J$ are given by
\begin{equation}
\label{eqn003}
\vec{F} = \mbox{diag}\left[\frac{R_2-R_1}{R_2},\,\frac{R_2-R_1}{R_2}\frac{r}{r-R_1},\,1\right],
\quad
J = \frac{(R_2-R_1)^2r}{R_2^2(r-R_1)}.
\end{equation}

Then, Navier equations (\ref{eqn001}) transform into 
\begin{eqnarray}
\nonumber
\frac{\lambda\!+\!2\mu}{r}\left[(r-R_1)\frac{\partial^2u_r}{\partial r^2}\!+\!\frac{\partial u_r}{\partial r}\!-\!\frac{u_r}{r-R_1}\right]\!+\!
\frac{\lambda\!+\!\mu}{r}\left[\frac{\partial^2u_\theta}{\partial r\partial \theta}+\frac{R_2(r-R_1)}{R_2-R_1}\frac{\partial^2u_z}{\partial r\partial z}\right]\\
\nonumber
-\frac{\lambda+3\mu}{r(r-R_1)}\frac{\partial u_\theta}{\partial \theta}
+\frac{\mu}{r}\left[\frac{1}{r-R_1}\frac{\partial^2u_r}{\partial \theta^2}+
\frac{R_2^2(r-R_1)}{(R_2-R_1)^2}\frac{\partial^2u_r}{\partial z^2}\right]=\hat \rho \frac{\partial^2u_r}{\partial t^2}, \\[4 mm]
\nonumber
\frac{\lambda\!+\!2\mu}{r(r-R_1)}\frac{\partial^2u_\theta}{\partial \theta^2}\!+\!
\frac{\lambda\!+\!3\mu}{r(r-R_1)}\frac{\partial u_r}{\partial \theta}\!+\!
\frac{\lambda\!+\!\mu}{r}\left(\frac{\partial^2u_r}{\partial r\partial \theta}+\frac{R_2}{R_2-R_1}\frac{\partial^2u_z}{\partial \theta\partial z}\right)\\
\nonumber
+\frac{\mu}{r}\left[
(r-R_1)\frac{\partial^2u_\theta}{\partial r^2}\!+\!\frac{\partial u_\theta}{\partial r}\!-\!\frac{u_\theta}{r-R_1}+
\frac{R_2^2(r-R_1)}{(R_2-R_1)^2}\frac{\partial^2 u_\theta}{\partial z^2}\right]
=\hat \rho \frac{\partial^2u_\theta}{\partial t^2}, \\[4 mm]
\nonumber
(\lambda\!+\!2\mu)\frac{R_2^2(r-R_1)}{(R_2-R_1)^2r}\frac{\partial^2u_z}{\partial z^2}\!+\!
\frac{\lambda\!+\!\mu}{r}\frac{R_2}{R_2-R_1}\left[(r-R_1)\frac{\partial^2u_r}{\partial r\partial z}+\frac{\partial^2u_\theta}{\partial \theta\partial z}+\frac{\partial u_r}{\partial z}\right]\\
+\frac{\mu}{r}\left[(r-R_1)\frac{\partial^2u_z}{\partial r^2}+\frac{1}{r-R_1}\frac{\partial^2 u_z}{\partial \theta^2}+\frac{\partial u_z}{\partial r}\right]
=\hat \rho \frac{\partial^2 u_z}{\partial t^2},
\label{eqn004}
\end{eqnarray}
where $\hat\rho=\rho/J$ and $R_1\le r \le R_2$.
These correspond to an extension to the three dimensional case of the equations given in \cite{brun2009}. Note that the identity gauge has been considered, i.e.  ${\bf u(\vect{x},t)}={\bf U(\vec{X},t)}$ where ${\bf u}=(u_r,u_\theta,u_z)$ and ${\bf x}=(r,\theta,z)$.

Equations (\ref{eqn004}) are accompanied by transformed boundary conditions on the upper and lower external surfaces $z=\pm h/2$:
\begin{eqnarray}
\label{eqn005}
\nonumber
\mu\left(\frac{\partial u_r}{\partial z}+\frac{R_2-R_1}{R_2}\frac{\partial u_z}{\partial r}\right)=0, &\\
\nonumber
\mu\left(\frac{R_2-R_1}{R_2}\frac{1}{r-R_1}\frac{\partial u_z}{\partial \theta}+\frac{\partial u_\theta}{\partial z}\right)=0,& \quad R_1\le r \le R_2\\
\lambda\frac{R_2-R_1}{R_2}\left(\frac{\partial u_r}{\partial r}+\frac{1}{r-R_1}\frac{\partial u_\theta}{\partial \theta}+\frac{u_r}{r-R_1}\right)+(\lambda+2\mu)\frac{\partial u_z}{\partial z}=0.&
\end{eqnarray}

\subsection{Asymptotic model}
In order to obtain the Kirchhoff plate model directly from the transformed equations of elasticity (\ref{eqn004}) and (\ref{eqn005}) an asymptotic procedure for elliptic operators in thin domains is developed \cite{MovMov1995,KozMazMov1999}. We introduce the scaled spatial variable $\xi=z/\epsilon$,
$\epsilon\ll 1$, and we also assume that the transverse displacement component depends on the scaled time variable $T=\epsilon\, t$. Then, we consider the following asymptotic approximation for the displacement vector ${\bf u}$
\begin{equation}
\label{eqn006}
{\bf u} \approx \sum_{k=0}^\infty \epsilon^k \left\{ \epsilon^{-4} \sum_{q=0}^3 \epsilon^q {\bf v}^{(q)}+  \epsilon^{-2} \sum_{q=0}^1 \epsilon^q {\mathcal V}^{(q)}+{\bf W}^{(k)}\right\},
\end{equation}
where ${\bf v}^{(q)}=(v_r^{(q)},v_\theta^{(q)},v_\xi^{(q)})$, $q = 0,1,2,3$, are functions of  $(r,\theta,\xi, T)$ 
and ${\mathcal V}^{(q)}=({\mathcal V}_r^{(q)},{\mathcal V}_\theta^{(q)},{\mathcal V}_\xi^{(q)})$,  $q = 0,1$, are functions of  $(r,\theta,\xi, t)$. The two finite sums on the right-hand side of equation (\ref{eqn006}) provide the solvability condition for ${\bf W}^{(0)}=(W^{(0)}_r,W^{(0)}_\theta,W^{(0)}_\xi)$ after substitution of the equation of motion (\ref{eqn004}) and boundary conditions (\ref{eqn005}). The solvability condition for $W^{(0)}_\xi$ constitutes a well-posed problem for the transverse displacement field $v_\xi^{(0)}$ describing the flexural behavior of a thin plate. 
The solvability conditions for $W^{(0)}_r$ and $W^{(0)}_\theta$ constitute a well-posed problem for the in-plane displacement field ${\mathcal V}_r^{(0)}$ and ${\mathcal V}_\theta^{(0)}$ describing the behavior of a thin shell.  Here interest is in the description of the plate model and we restrict attention to the asymptotic procedure for  $v_\xi^{(0)}$ which will be indicated by $v$ for ease of notation.

After the introduction of the scaled variable $\xi$, the equation of motion (\ref{eqn004}) and boundary conditions (\ref{eqn005}) can be expressed in the form
\begin{eqnarray}
\label{eqn006b}
\left(\frac{1}{\epsilon^2}\mathcal{L}_0+\frac{1}{\epsilon}\mathcal{L}_1+\mathcal{L}_2 \right){\bf u}=\epsilon^2\hat\rho\frac{\partial^2 {\bf u}}{\partial T^2}
\end{eqnarray}
in $(R_1\le r\le R_2, 0\leq \theta <2\pi,-H/2\leq \xi\leq H/2)$, with $H=h/\epsilon$, and
\begin{eqnarray}
\label{eqn007}
\left(\frac{1}{\epsilon}{\Sigma_0}+{\Sigma_1} \right){\bf u}={\bf 0}
\end{eqnarray} 
on $(R_1\le r\le R_2, 0\leq \theta <2\pi,\xi=\pm H/2)$.

In equation (\ref{eqn006b}) 
\begin{eqnarray}
\nonumber
\label{eqn008}
\mathcal{L}_0=\frac{r-R_1}{r}\left(\frac{R_2}{R_2-R_1}\right)^2 
\left( \begin{array}{ccc}
\mu\frac{\partial^2}{\partial \xi^2} & 0 & 0 \\
0 & \mu\frac{\partial^2}{\partial \xi^2} & 0 \\
0 & 0 &  (\lambda+2\mu)\frac{\partial^2}{\partial \xi^2} \\
\end{array}\right), \\ [4 mm]
\nonumber
\mathcal{L}_1=\frac{R_2}{R_2-R_1}\frac{\lambda+\mu}{r} 
\left( \begin{array}{ccc}
0 & 0 & (r-R_1)\frac{\partial^2}{\partial \xi \partial r} \\
0 & 0 & \frac{\partial^2}{\partial \xi \partial \theta} \\
\frac{\partial}{\partial r}\left[(r-R_1)\frac{\partial}{\partial \xi}\right] & \frac{\partial^2}{\partial \xi \partial \theta} & 0 \\
\end{array}\right),\\[4 mm]
\mathcal{L}_2=\frac{1}{r(r-R_1)} 
\left( \begin{array}{ccc}
\mathcal{L}_2^{[11]} & \mathcal{L}_2^{[12]} & 0 \\
\mathcal{L}_2^{[21]} & \mathcal{L}_2^{[22]} & 0 \\
0 & 0 &  \mu(r-R_1)^2\hat\nabla^2_{R_1} \\
\end{array}\right),
\end{eqnarray}
where
\begin{eqnarray}
\nonumber
\mathcal{L}_2^{[11]}=\mu\frac{\partial^2}{\partial \theta^2} + (\lambda+2\mu)\left\{(r-R_1)\frac{\partial}{\partial r}\left[(r-R_1)\frac{\partial}{\partial r}\right]-1\right\},\\
\nonumber
\mathcal{L_2^{[12]}}=-(\lambda+3\mu)\frac{\partial}{\partial \theta} + (\lambda+\mu)(r-R_1)\frac{\partial^2}{\partial r\partial \theta},\\
\nonumber
\mathcal{L}_2^{[21]}=(\lambda+3\mu)\frac{\partial}{\partial \theta} + (\lambda+\mu)(r-R_1)\frac{\partial^2}{\partial r\partial \theta},\\
\mathcal{L}_2^{[22]}=(\lambda+2\mu)\frac{\partial^2}{\partial \theta^2} + \mu\left\{(r-R_1)\frac{\partial}{\partial r}\left[(r-R_1)\frac{\partial}{\partial r}\right]-1\right\} 
\end{eqnarray}
and the differential operator $\hat\nabla^2_{R_1}$ is defined in equation (\ref{eq030a}).

For equation (\ref{eqn007}) 
\begin{eqnarray}
\nonumber
\label{eqn009}
\Sigma_0=
\left( \begin{array}{ccc}
\mu\frac{\partial}{\partial \xi} & 0 & 0 \\
0 & \mu\frac{\partial}{\partial \xi} & 0 \\
0 & 0 &  (\lambda+2\mu)\frac{\partial}{\partial \xi} \\
\end{array}\right), \\ [4 mm]
\Sigma_1=\frac{R_2-R_1}{R_2}
\left( \begin{array}{ccc}
0 & 0 & \mu\frac{\partial}{\partial r} \\[2 mm]
0 & 0 & \mu\frac{1}{r-R_1}\frac{\partial}{\partial \theta} \\
\lambda\left(\frac{\partial}{\partial r}+\frac{1}{r-R_1}\right) & \frac{\lambda}{r-R_1}\frac{\partial}{\partial \theta} & 0 \\
\end{array}\right).
\end{eqnarray}

\subsubsection{Hierarchical system of equations}
A hierarchical system of equations is obtained by substituting the asymptotic representation (\ref{eqn006}) into transformed field equations (\ref{eqn004}) complemented by the transformed boundary conditions (\ref{eqn005}). 
 
To leading order, the equations
\begin{equation}
\mathcal{L}_0{\bf v^{(0)}}= {\bf 0}
\label{eqn010}
\end{equation}
with boundary conditions
\begin{equation}
\Sigma_0{\bf v^{(0)}}={\bf 0}
\label{eqn011}
\end{equation}
are satisfied by
\begin{equation}
{\bf v^{(0)}}=
\left( \begin{array}{c} 
0 \\
0 \\
v(r,\theta,T) \\
\end{array}\right),
\label{eqn012}
\end{equation}
where $v$ does not depend on $\xi$ and the solvability conditions are automatically satisfied.

To the next order, the field equations 
\begin{equation}
\mathcal{L}_0{\bf v}^{(1)}+\mathcal{L}_1{\bf v}^{(0)}= {\bf 0}
\label{eqn013}
\end{equation}
and boundary conditions
\begin{equation}
\Sigma_0{\bf v}^{(1)}+\Sigma_1{\bf v}^{(0)}={\bf 0}
\label{eqn014}
\end{equation}
admit the solution
\begin{equation}
{\bf v^{(1)}}=
-\frac{R_2-R_1}{R_2}\left( 
\begin{array}{c} 
\frac{\partial v}{\partial r} \\[2 mm]
\frac{1}{r-R_1}\frac{\partial v}{\partial \theta} \\[2 mm]
0 \\
\end{array}\right) \xi.
\label{eqn015}
\end{equation}
Note that ${\bf v^{(1)}}$, and ${\bf v^{(2)}}$ and ${\bf v^{(3)}}$ in the following, are complemented by the normalisation condition of zero average along the thickness.

Next, the field equations 
\begin{equation}
\mathcal{L}_0{\bf v}^{(2)}+\mathcal{L}_1{\bf v}^{(1)}+\mathcal{L}_2{\bf v}^{(0)}= {\bf 0}
\label{eqn016}
\end{equation}
and boundary conditions
\begin{equation}
\Sigma_0{\bf v}^{(2)}+\Sigma_1{\bf v^{(1)}}={\bf 0}
\label{eqn017}
\end{equation}
give
\begin{equation}
{\bf v^{(2)}}=
\left(\frac{R_2-R_1}{R_2}\right)^2\left( 
\begin{array}{c} 
0 \\
0 \\
\hat\nabla^2_{R_1}v
\end{array}\right)\frac{\lambda}{\lambda+2\mu} \left(\frac{\xi^2}{2}-\frac{H^2}{24}
\right).
\label{eqn018}
\end{equation}

For the following order, we have the equations
\begin{equation}
\mathcal{L}_0{\bf v}^{(3)}+\mathcal{L}_1{\bf v}^{(2)}+\mathcal{L}_2{\bf v}^{(1)}= {\bf 0}
\label{eqn019}
\end{equation}
accompanied by the boundary conditions 
\begin{equation}
\Sigma_0{\bf v^{(3)}}+\Sigma_1{\bf v}^{(2)}={\bf 0}
\label{eqn020}
\end{equation}
and the corresponding solution is
\begin{equation}
{\bf v^{(3)}}=
\left(\frac{R_2-R_1}{R_2}\right)^3\left( 
\begin{array}{c} 
\frac{\partial}{\partial r} \left(\hat\nabla^2_{R_1} v\right)\\[2 mm]
\frac{1}{r-R_1}\frac{\partial}{\partial \theta} \left(\hat\nabla^2_{R_1} v\right) \\[2 mm]
0 \\
\end{array}\right) \frac{(3\lambda+4\mu)\xi^3/6-(11\lambda+12\mu)\xi H^2/24}{\lambda+2\mu}.
\label{eqn021}
\end{equation}
is defined in equation (\ref{eq030a}).

Finally, the vector function ${\bf W}^{(0)}$ satisfies the equation
\begin{equation}
\mathcal{L}_0{\bf W}^{(0)}+\mathcal{L}_1{\bf v}^{(3)}+\mathcal{L}_2{\bf v}^{(2)}=\hat\rho \frac{\partial^2{\bf v^{(0)}}}{\partial T^2}
\label{eqn022}
\end{equation}
together with the boundary conditions
\begin{equation}
\Sigma_0{\bf W^{(0)}}+\Sigma_1{\bf v}^{(3)}={\bf 0}.
\label{eqn023}
\end{equation}

In particular, $W^{(0)}_\xi$ solves the problem 
\begin{equation}
\frac{\partial^2 W^{(0)}_\xi}{\partial \xi^2}=-\left(A_1\frac{\xi^2}{2}-A_2\frac{H^2}{24}\right)\hat\nabla^2_{R_1}(\hat\nabla^2_{R_1} v)+\frac{\rho}{\lambda+2\mu} \frac{\partial^2 v}{\partial T^2}=F, \,\,\,\,\,\, |\xi|<H/2
\label{eqn027}
\end{equation}
with
\begin{equation}
A_1=\left(\frac{R_2-R_1}{R_2}\right)^4\frac{3\lambda+2\mu}{\lambda+2\mu}, \quad
A_2=\left(\frac{R_2-R_1}{R_2}\right)^4\frac{11\lambda^2+24\lambda\mu+12\mu^2}{(\lambda+2\mu)^2},
\label{eqn028}
\end{equation}
subjected to boundary conditions
\begin{equation}
\left.\frac{\partial W^{(0)}_\xi}{\partial \xi}\right|_{\xi=\pm \frac{H}{2}}=
\pm\left(\frac{R_2-R_1}{R_2}\right)^4\frac{\lambda\left(\lambda+\mu\right)}{\left(\lambda+2\mu\right)^2}\frac{H^3}{6}\hat\nabla^2_{R_1}(\hat\nabla^2_{R_1} v)
=p_\pm.
\label{eqn026}
\end{equation}
The solvability condition
\begin{equation}
\int_{-H/2}^{H/2}Fd\xi=p_+-p_-
\label{eqn025}
\end{equation}
is the fourth-order differential equation
\begin{equation}
\left(\frac{R_2-R_1}{R_2}\right)^4\frac{E h^3}{12(1-\nu^2)}\hat\nabla^2_{R_1}(\hat\nabla^2_{R_1}v) + \rho h \frac{\partial^2 v}{\partial t^2}=0,
\label{eqn024}
\end{equation}
where $E$ is the Young's modulus, $\nu$ the Poisson's ratio and $H$ and $T$ have been replaced by $h/\epsilon$ and $\epsilon\,t$, respectively. 

Identifying the flexural rigidity $D^{(0)}$ with the coefficient $E h^3/(12(1-\nu^2))$ it is straightforward to check that equations (\ref{eq030}) and  (\ref{eqn024}), restricted to time-harmonic regime, are the same.

\section{Conclusions}

There are different ways of reducing the shadow generated by a scatterer. In particular, an elementary example where a ``heavy'' inclusion is surrounded by a ``lighter'' isotropic coating was discussed in \cite{FarCheBagEnoGueAlu2014}. That model requires the average mass density of the inclusion and coating together to be the same as the mass density of the ambient matrix. Such examples have been known for more than a century (see, for example, \cite{Voigt1889}). It is important to mention that a combination of a heavy inclusion and a lighter coating cannot be associated with an ``invisibility cloak'', but instead can be used to reduce the monopole source term in the asymptotics at infinity.

In \cite{stenger2012} the use of micro-structured material for cloaking represented a substantial advance. That work has demonstrated that cloaking of a flexural wave is possible, although such a  cloaking approximation is frequency dependent. In the present paper, we have provided a full theoretical background for such an approximation and have also discussed the range of its applicability.

Furthermore, by referring to pre-stressed elastic  plates, we have resolved a long-standing problem of creating an exact cloak for flexural  waves.  For the cloaking region obtained as a result of a ``push-out'' radially symmetric transformation, we have identified  a full set of parameters, including pre-stress and in-plane body forces. 
In the case when pre-stress and body forces are not included in the model, an approximation of the cloak has been developed for $R_1/R_2\ll1$ and within the frequency range when $\beta R_1\ll 1$.
The illustrative numerical computations show excellent agreement with the prediction of the theoretical model and the existing experimental results. 

The transformed equations of three-dimensional vector elasticity were analysed asymptotically for a thin solid. The resulting lower-dimensional model agrees fully with the outcome of the direct application of the radial ``push-out'' transformation to the equation of 
motion of a Kirchhoff plate. 
It is also noted that `transformed' material in three-dimensional elasticity has non-symmetric constitutive relations, as outlined in \cite{brun2009}, but the lower-dimensional model for the plate does not have such a feature. The physical nature of the reduced model is fully explained, with the introduction of pre-stress and in-plane body forces, which have been identified in explicit closed form.
Implementation of the proposed model could lead to a new generation of lightweight and highly-efficient structured shields and filtering devices.

\section*{Acknowledgment}
{\small
A.B.M. and N.V.M. acknowledge the financial support of the European Community's Seven Framework Programme under contract numbers PIAP-GA-2011-286110-INTERCER2 and  PIAPP-GA-284544-PARM-2.
M.B. acknowledges the financial support of the European Community's Seven Framework Programme under contract number PIEF-GA-2011-302357-DYNAMETA and of  Regione Autonoma della Sardegna (LR7 2010, grant `M4' CRP-27585).
D.J.C. acknowledges the financial support of EPSRC in the form of a Doctoral Prize Fellowship and grant EP/J009636/1.
}

\appendix
\section{Appendix: Kirchhoff plate versus Mindlin model  in the numerical implementation}
\label{App1}

The commercial finite element software Comsol Multiphysics\textsuperscript{\textregistered} was used to produce the numerical simulation presented in \S\ref{sec:numerics}.
In Comsol, plates are implemented using the Mindlin-Reissner model \cite{Reissner1945,Mindlin1951} accounting for the shear deformation through the thickness of the plate.
The equation governing the flexural displacement of a homogeneous isotropic Mindlin plate is
\begin{equation}
\left(\nabla^2 + \frac{\rho}{G}\frac{\partial^2}{\partial t^2}\right)\left(D\nabla^2 - \frac{\rho h^3}{12}\frac{\partial^2}{\partial t^2}\right)w + \rho h\frac{\partial^2 w}{\partial t^2} = 0,
\label{eq:gov-mindlin}
\end{equation}
where $\rho$ is the density, $G$ is the shear modulus, $D$ is the flexural rigidity, $h$ is the thickness of the plate, and $w$ is the flexural displacement.
On the other hand, the corresponding equation for a Kirchhoff-Love plate is $D\nabla^4 w + \rho h\ddot{w} = 0$.
Assuming that $D\sim\mathcal{O}(1)$ and $\rho h\sim\mathcal{O}(1)$ we observe that equation \eqref{eq:gov-mindlin}, approximately, reduces to the governing equation for the Kirchhoff-Love plate provided that $\rho/G \ll 1$ and $\rho h^3 \ll1$.
Thus, using a judicious choice of parameters, the additional terms in \eqref{eq:gov-mindlin} introduced by accounting for the variation of the shear deformation through the thickness of the plate can be neglected.
In this way, the Kichhoff-Love plate equation may be simulated using finite element models built using the Mindlin-Reissner Comsol package.
A more detailed comparison of the dynamics of Kirchhoff-Love and Mindlin plates can be found in~\cite{movchan2011}.

\begin{figure}[!htcb]
\centering
\begin{subfigure}[c]{0.3\linewidth}
\includegraphics[width=\linewidth]{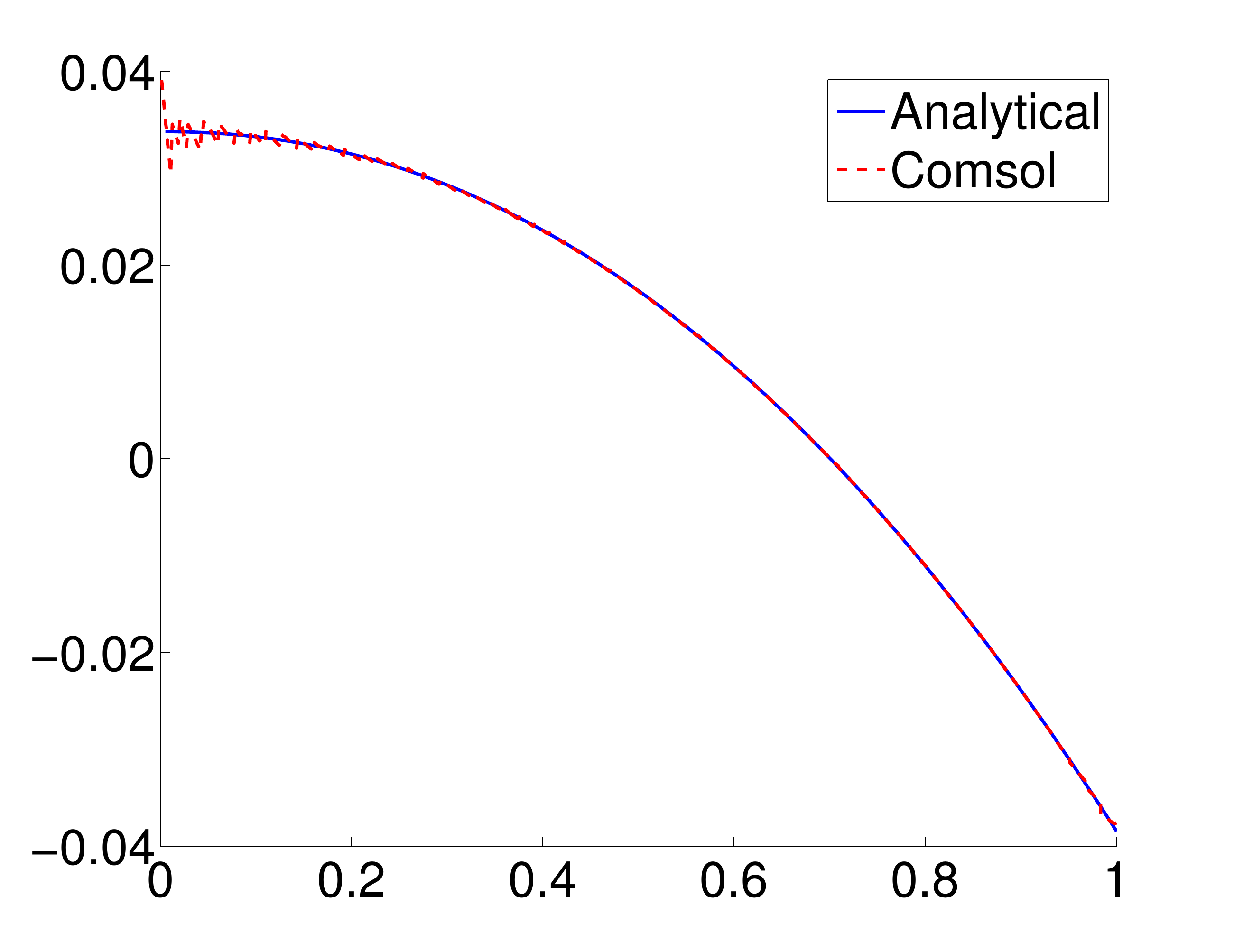}
\caption{$w$}
\end{subfigure}
\quad
\begin{subfigure}[c]{0.3\linewidth}
\includegraphics[width=\linewidth]{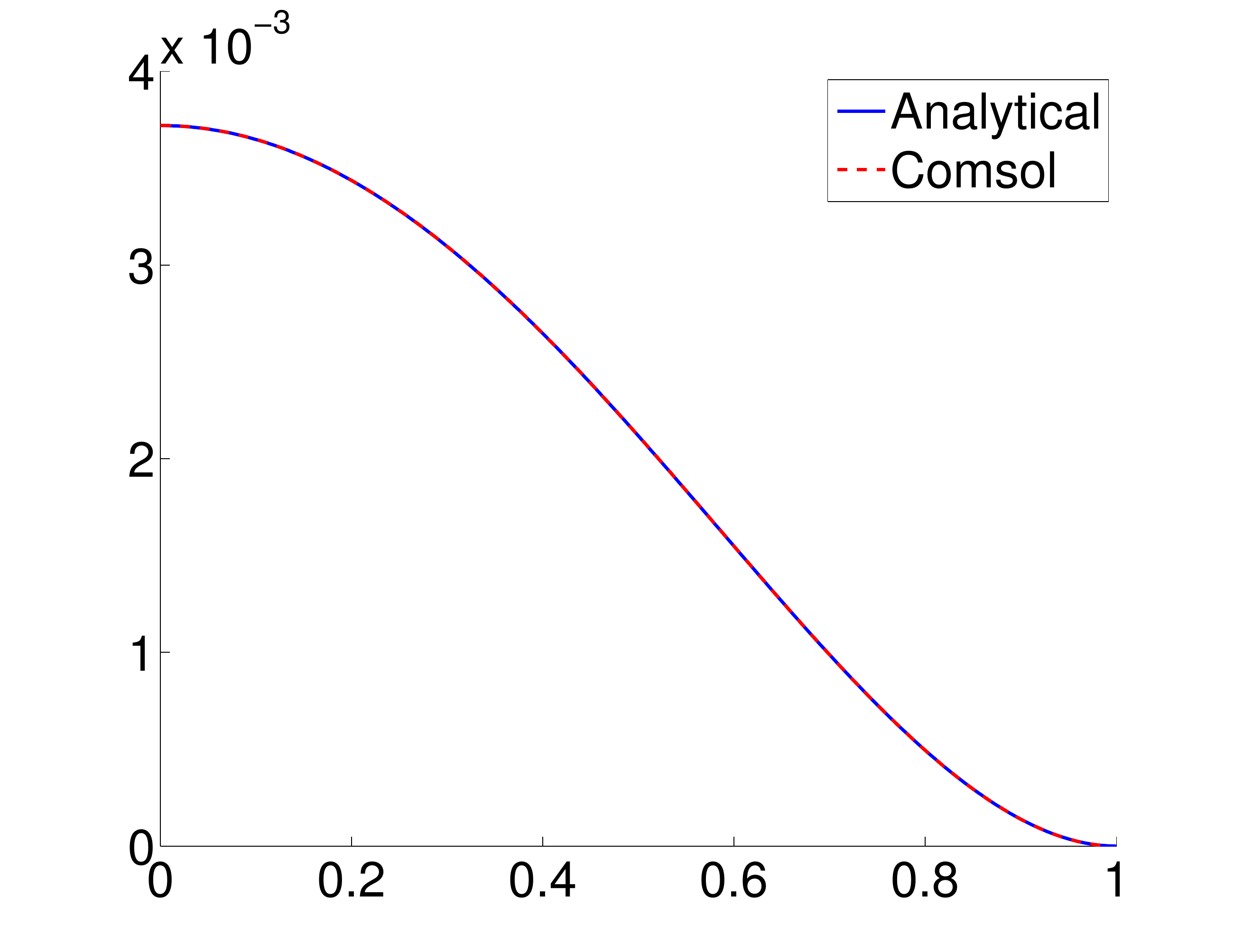}
\caption{$M_{r}$}
\end{subfigure}
\quad
\begin{subfigure}[c]{0.3\linewidth}
\includegraphics[width=\linewidth]{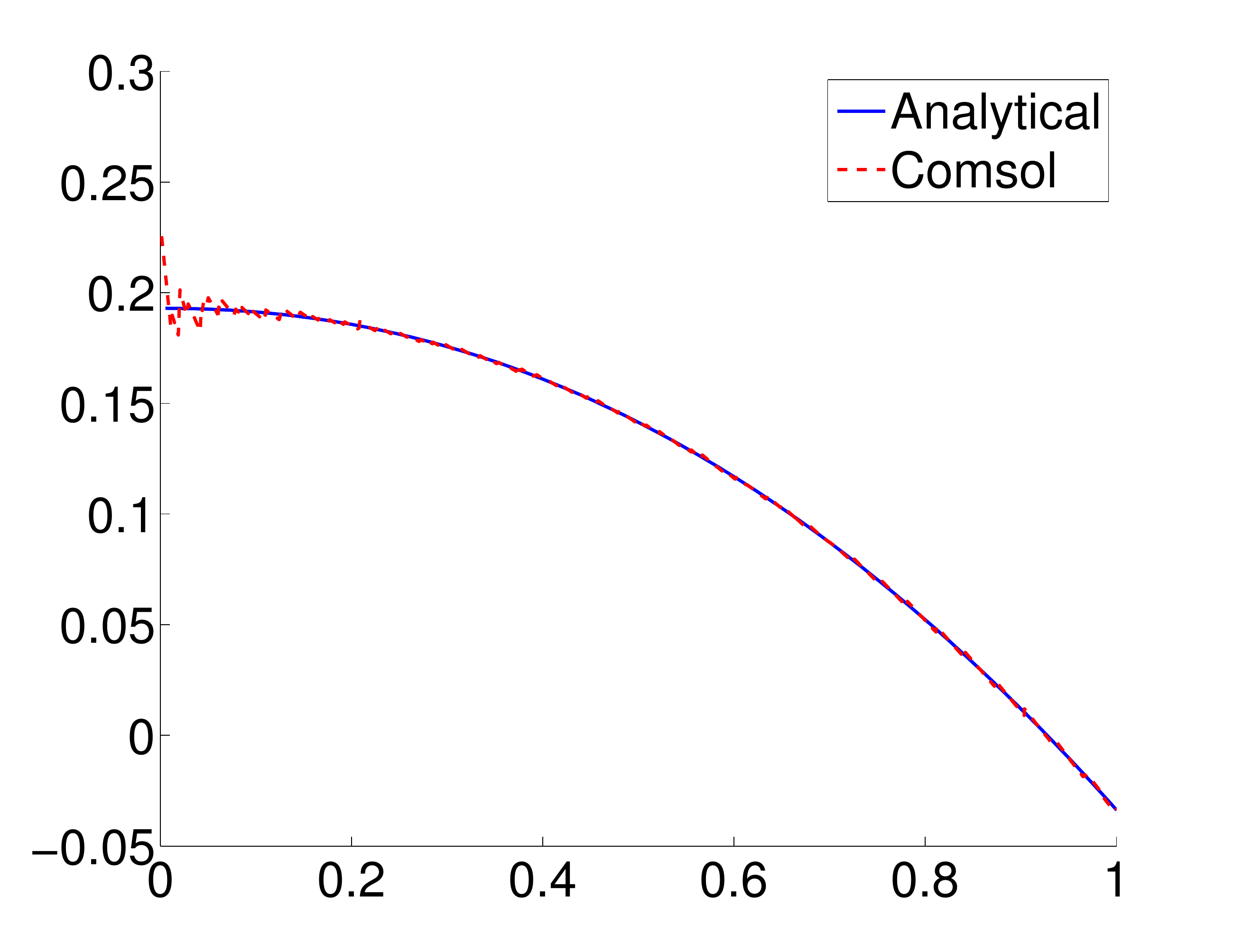}
\caption{$M_{\theta}$}
\end{subfigure}
\caption{\label{fig:verifcation}
The analytical solution for a radially orthotropic Kirchhoff-Love plates (solid blue line) compared with the numerical solution for Mindlin plates (dashed red line).
The numerical values used were $D_\theta = 12.3$, $D_r = 1.25$, $\nu_\theta = 0.876$, $\nu_r = 0.0890$, $h=0.001$ and $G\approx 10^{10}$.}
\end{figure}

For the purpose of the numerical simulations presented in \S\ref{sec:numerics}, the shear modulus was chosen as $G\approx 10^{10}$ and $h = 1\times10^{-3}$, while all other parameters were chosen as unity.
In order to verify this approach, Comsol's Mindlin plate model was used to compute a static verification model.
In particular, we consider the Green's function for a homogeneous radially orthotropic circular plate of radius $R_2$ with clamped boundaries.
This problem was considered for Kirchhoff-Love plates in \cite{lekhnitskii1968} $\S 82$ (see also \cite{Carrier1944}) and has the following analytical solution
\begin{eqnarray}
\nonumber
w(R)=\frac{R_2^2}{4\pi D_R(1-\eta^2)(1+\eta)}\left[1-\eta+(1+\eta)\left(\frac{R}{R_2}\right)^2-2\left(\frac{R}{R_2}\right)^{1+\eta}\right],\\
\nonumber
M_R=\frac{1}{2\pi(1-\eta^2)}\left[(\eta+\nu_\Theta)\left(\frac{R}{R_2}\right)^{\eta-1}-(1+\nu_\Theta)\right],\\
\nonumber
M_\Theta=\frac{\eta^2}{2\pi(1-\eta^2)}\left[(1+\eta\nu_R)\left(\frac{R}{R_2}\right)^{\eta-1}-(1+\nu_R)\right],\\
M_{R\Theta}=0.
\label{App001}
\end{eqnarray}
where $\eta=\sqrt{D_\Theta/D_R}$.
Figure \ref{fig:verifcation} shows the agreement between the analytical solution for Kirchhoff-Love plates and the numerical solution produced using Comsol for Mindlin plates.

\end{document}